\documentclass[%
 aip,
 amsmath,amssymb,
 reprint,%
]{revtex4-1}

\usepackage{graphicx}
\usepackage{dcolumn}
\usepackage{bm}
\usepackage{color}

\usepackage[utf8]{inputenc}
\usepackage[T1]{fontenc}
\usepackage{mathptmx}
\usepackage{gensymb}
\usepackage{float}
\usepackage{simplewick}
\usepackage{amsmath}
\usepackage{comment}

\begin{document}

\preprint{AIP/123-QED}

\title{Non-covalent interactions between molecular dimers (S66) in electric fields}
\author{Max Schwilk}
\thanks{Contributed equally}
\affiliation{Faculty of Physics, 
University of Vienna, 
Kolingasse 14-16, 
1090 Vienna, 
Austria}
\affiliation{Institute of Physical Chemistry and National Center for Computational Design and Discovery of Novel Materials (MARVEL), Department of Chemistry, University of Basel, Klingelbergstrasse 80, CH-4056 Basel, Switzerland}
\author{P\'{a}l D. Mezei}
\thanks{Contributed equally}
\affiliation{Institute of Physical Chemistry and National Center for Computational Design and Discovery of Novel Materials (MARVEL), Department of Chemistry, University of Basel, Klingelbergstrasse 80, CH-4056 Basel, Switzerland}
\author{Diana N. Tahchieva}
\affiliation{Institute of Physical Chemistry and National Center for Computational Design and Discovery of Novel Materials (MARVEL), Department of Chemistry, University of Basel, Klingelbergstrasse 80, CH-4056 Basel, Switzerland}
\author{O. Anatole von Lilienfeld}
\email{anatole.vonlilienfeld@univie.ac.at}
\affiliation{Faculty of Physics, 
University of Vienna, 
Kolingasse 14-16, 
1090 Vienna, 
Austria}
\affiliation{Institute of Physical Chemistry and National Center for Computational Design and Discovery of Novel Materials (MARVEL), Department of Chemistry, University of Basel, Klingelbergstrasse 80, CH-4056 Basel, Switzerland}

\date{\today}

\begin{abstract}
Fine tuning and microscopic control of van der Waals interactions through oriented external electric fields (OEEF) 
mandates an accurate and systematic understanding of intermolecular response properties. 
Having taken exploratory steps into this direction, we present a systematic study of interaction induced dipole electric properties
of all molecular dimers in the S66 set, relying on CCSD(T)-F12b/aug-cc-pVDZ-F12 as reference level of theory. 
For field strengths up to $\approx$5 GV m$^{-1}$ the interaction induced electric response beyond second order is found to be insignificant.
Large interaction dipole moments (i.e.~dipole moment changes due to van der Waals binding) 
are observed in the case of hydrogen bonding oriented along the intermolecular axis, 
and mostly small interaction dipole moments are found in dimers bonded by $\pi$-stacking or London dispersion. 
The interaction polarizabilities (i.e.~polarizability changes due to van der Waals binding) were generally found to be small but always with a positive-valued principal component approximately aligned with the intermolecular axis, and two other negative-valued components. 
Energy decompositions according to symmetry adapted perturbation theory (SAPT0/jun-cc-pVDZ)
suggest that electrostatics dominates the interaction dipole moment, 
with exchange and induction contributing on a smaller scale, and with dispersion having the smallest effect.
First-order SAPT0 decomposition into monomer-resolved contributions enables us to 
establish a quantitative link between electric properties of monomers and dimers,
which is found to be in qualitative agreement with the coupled cluster reference method. 
Using the aug-cc-pVQZ basis and non-empirical PBE semilocal exchange-correlation kernels, 
we also assess how density functional approximations in the nonlocal exchange and correlation parts affect the predictive accuracy:
While dRPA@PBE0 based predictions are in excellent overall agreement with coupled cluster results, 
the computationally more affordable LC-$\omega$PBE0-D3 level of theory also yields reliable results with relative errors below 5\%. 
PBE alone, even when dispersion corrected, produces larger errors 
in interaction dipole moments ($\approx$ 10\%) and polarizabilities ($\approx$ 20\%). 
We also resolve the mutual impact of the three dimensions of the OEEF, 
and we present a discussion of the intermolecular distance dependence of the perturbations.
\end{abstract}

\maketitle

\section{Introduction}

Modulating chemical interactions of molecules with an external electric field is a rather intuitive concept that nevertheless has not yet found its way to widespread synthetic applications.
An external electric field can be present as a result of a crystal lattice or a solvent environment (with field strengths of up to 8 GV m$^{-1}$),\cite{spackman2007use,fried2017electric} and it also occurs between the tip and the sample in scanning tunneling microscopy.\cite{hong1998electron}
It was only in the 1970s that the capability of external electric fields to catalyze chemical reactions was demonstrated by exploring the ``intramolecular salt effect''.\cite{pocker1970electro,SMITH19912617}
The experimental work also triggered computational efforts on probing the chemical modulation capabilities of the oriented external electric field (OEEF), such as accelerating proton transfer reactions,\cite{arabi2011effects} catalyzing the Diels-Alder reaction,\cite{meir2010oriented} or controlling reaction selectivity.\cite{shaik2004external}
Experimental progress included the activation of enzymes\cite{thuren1987triggering} and the reversible isomerization by the OEEF at the tip of a scanning tunneling microscope.\cite{alemani2006electric}
In 2016, Aragon{\`e}s \emph{et al.}\cite{aragones2016} succeeded in using the OEEF at the tip of the scanning tunneling microscope to catalyze the Diels-Alders reaction.
This breakthrough led to a recent surge in efforts to unveil the full potential of OEEFs as a chemical tool.
Impressive progress has been made due to synergies of experimental and theoretical efforts.\cite{shaik2016oriented,Nejad2014,shaik2020,kaestner2020} 

Strong electric fields can also occur as an intramolecular effect.
An internal electric field can arise in the cavity of frustrated Lewis pairs\cite{grimme2010mechanism}. 
An internal electric field of mainly dipolar character is generated at an active site of enzymes if two oppositely charged residues are in its close vicinity.\cite{lehle2005probing,suydam2006electric}
In proteins, induced internal electric fields can be measured by the vibrational Stark effect.\cite{fried2015measuring} 
An internal electric field can also be adjusted by point mutations in proteins (such as replacing residues in the amino acid sequence) for example to accelerate photosynthetic reactions,\cite{gopher1985effect,popovic1986electric,franzen1990electric} or to alter the enzyme activity.\cite{moroney1984effect} 

Noncovalent interactions play an important role for understanding supra-molecular processes, developing new synthesis pathways,\cite{knowles2010attractive} designing molecular recognition processes,\cite{metrangolo2008halogen}, or engineering materials.\cite{zhang2003fabrication}
For example, within a recent study, molecular dynamics simulations revealed that a novel paracetamol polymorph can be created in the presence of a strong electric field (1.5 GV m$^{-1}$).\cite{parks2017molecular}
Or, Kleshchonok and Tkatchenko studied intermolecular interactions in the presence of point charges in biologically 
relevant noncovalent dimers,\cite{kleshchonok2018tailoring}
obtaining field induced dispersion effects amounting to up to 35\% of the total interaction energy.
Even more surprisingly, they found that positive and negative point charges in the vicinity 
of noncovalent dimers may weaken or strengthen the dispersion interaction.
Similarly, non-trivial static effects can also emerge when it comes to the screening of dispersion in layered systems~\cite{ambrosetti2018hidden}.

Koz{\l}owska \emph{et al.}\cite{kozlowska} computed interaction induced dipole moments of endohedral small polar molecules in carbon nanotubes of varying diameter with the PNO-LCCSD-F12\cite{Schwilk2017JCTC,Ma2017JCTC_1} local coupled cluster method and compared various density functional approximations (DFAs), addressing the impact of exchange-correlation kernels, orbital relaxation, and dispersion treatments.
Their results underline the sensitivity of DFAs with respect to the dispersion treatment for electric properties. 
In a follow-up study, DFAs were benchmarked for interaction-induced electric properties of small hydrogen-bonded dimers confined by a simple quadratic potential.\cite{kozlowska2} 
It should be noted that, in general, interaction induced electric properties may also contain important contributions from nuclear relaxation and molecular vibration.\cite{dft_interaction_induced3}

In this work, we computationally quantify and examine the effect of a dipolar OEEF on noncovalent interaction energies for the S66 set with 66 noncovalent dimers of small and medium size molecules.\cite{s66,s66x8} 
The focus of our study lies on probing the capability of a dipolar OEEF to modulate the intermolecular interaction, on probing computationally more efficient methodologies to describe the effect, as well as on obtaining a qualitative understanding of the most relevant physical interactions.

\section{Methodology and computational details}

The equilibrium S66 noncovalent dimer geometries are taken from the literature,\cite{s66x8} and no further geometry optimization is applied.
For all S66 dimers, we calculate the field $\boldsymbol{F}$ dependent total potential energies of interaction $\Delta E$ according to eq \ref{inten}.

\begin{equation}\label{inten}\Delta E(\boldsymbol{F}) = E^{AB}(\boldsymbol{F})-E^{A}(\boldsymbol{F})-E^{B}(\boldsymbol{F})\end{equation}

In order to describe the change in the interaction energy induced by the applied OEEFs, we compute the interaction dipole moments ($\Delta \boldsymbol{\mu}$) and interaction polarizabilities ($\Delta \boldsymbol{\alpha}$) components $i,j\in\{x,y,z\}$ according to eqs \ref{intmu} and \ref{intalpha} by finite differentiation. The accuracy of this second order expansion is verified. 

\begin{equation}\label{intmu}\Delta \mu_{i}=-\frac{\partial \Delta E(\boldsymbol{F})}{\partial F_i} \Bigr|_{\boldsymbol{F}=\boldsymbol{0}}\end{equation}

\begin{equation}\label{intalpha}\Delta \alpha_{ij}=-\frac{\partial^2 \Delta E(\boldsymbol{F})}{\partial F_i \partial F_j} \Bigr|_{\boldsymbol{F}=\boldsymbol{0}}\end{equation}

An OEEF is applied along the intermolecular (\textit{z}) axis (going through the two monomer centers of mass and directed such that the \textit{z}-component of the sum of the two monomer dipole moments is positive) with values of 0.000 a.u., $\pm$0.001 a.u., $\pm$0.005 a.u., and $\pm$0.01 a.u. (0.01 a.u. $\approx$ 5.14 GV m$^{-1}$), covering a range of frequently encountered field strengths.
We also apply a three-dimensional grid of OEEF strengths $\{F_x,F_y,F_z\}$ with values of 0.00 a.u. and $\pm$0.01 a.u. in each dimension.

Furthermore, we analyze orientation and strength of the interaction dipole moment $\Delta\boldsymbol{\mu}$ as well as the interaction polarizability principal components $\boldsymbol{\nu}^r$, $r\in\{a,b,c\}$ as given in eq \ref{princip}.

\begin{equation}\label{princip}\begin{aligned}\left[\begin{matrix}v^{a}_x & v^{b}_x & v^{c}_x \\ v^{a}_y & v^{b}_y & v^{c}_y \\ v^{a}_z & v^{b}_z & v^{c}_z\end{matrix}\right]^{T} \left[\begin{matrix}\Delta\alpha_{xx} & \Delta\alpha_{xy} & \Delta\alpha_{xz} \\ \Delta\alpha_{yx} & \Delta\alpha_{yy} & \Delta\alpha_{yz} \\ \Delta\alpha_{zx} & \Delta\alpha_{zy} & \Delta\alpha_{zz}\end{matrix}\right] \left[\begin{matrix}v^{a}_x & v^{b}_x & v^{c}_x \\ v^{a}_y & v^{b}_y & v^{c}_y \\ v^{a}_z & v^{b}_z & v^{c}_z\end{matrix}\right] \\ = \left[\begin{matrix}\Delta\alpha_{aa} & 0 & 0 \\ 0 & \Delta\alpha_{bb} & 0 \\ 0 & 0 & \Delta\alpha_{cc}\end{matrix}\right]\end{aligned}\end{equation}

For the sake of completeness, we also present two simpler but more frequently used quantities in the supporting information (SI): the isotropic and anisotropic components of the interaction polarizability in the $(x,y,z)$ frame.
It has been shown that fourth order perturbation theory effects are important in the accurate description of field induced electric properties.\cite{dft_interaction_induced} 
We therefore choose the coupled cluster singles, doubles, and perturbative triples [CCSD(T)]\cite{mhg1989} 
method for reference. The coupled cluster reference calculations are performed with the Molpro quantum chemistry software\cite{MOLPRO_brief}).
Since the basis set convergence of CCSD(T) is fairly slow and comes with high computational overhead even for relatively small systems, we use the explicitly correlated F12b method\cite{knizia:2009} [3*A(fix,nox) variant as implemented in Molpro] 
with the aug-cc-pVDZ-F12\cite{Dunning1989,peterson2008systematically,nitai2017}
basis set and the aug-cc-pVTZ-F12 auxiliary basis set.\cite{nitai2017,Yousaf2008,Yousaf2009} 
This level of theory has shown to yield a reasonably small interaction energy error (0.048 kcal mol$^{-1}$ on average) at zero field strength for the benchmark S66 energies.\cite{s66x8}

We use the coupled cluster reference for a systematic assessment of different  density functional theory (DFT) approximations based on the semilocal nonempirical PBE functional.\cite{pbe}
Due to the rigorous roots in fundamental physics of the PBE functional we expect it to  cover short range exchange and correlation at least qualitatively correct in the context of electric response properties.\cite{dft_dipole_moment,dft_polarizability}
Approximations tested include the PBE-D3,\cite{pbe,dftd3,dftd3bj}, PBE-VV10,\cite{pbe,vv10} PBE0-D3,\cite{pbe0,dftd3,dftd3bj}, PBE0-VV10,\cite{pbe0,vv10} HSE06-D3,\cite{hse03,hse06,dftd3,dftd3bj} LC-$\omega$PBE-D3,\cite{lcwpbe,dftd3,dftd3bj} PBE0-2,\cite{pbe02}, dRPA75,\cite{drpa75}, dRPA@PBE, and dRPA@PBE0\cite{drpa_drccd,drpa_equivalence,drpa_implementation} functionals.
Hence, the numerical results can be expected to be fairly transparent regarding effects due to dispersion corrections, exact exchange mixing, and nonlocal correlation components, and encode the prospect that conclusions are mostly transferable to other density functional families.
All DFT computations are performed with the aug-cc-pVQZ basis set\cite{Dunning1989} in the MRCC quantum chemistry program suite.\cite{MRCC}

To gain further insights, we also employ symmetry-adapted perturbation theory (SAPT)\cite{jeziorski1994perturbation}
to decompose the interaction energy, dipole moment, and polarizability into physically meaningful contributions (such as electrostatic, exchange, induction, and dispersion).
In SAPT, the perturbed Hamiltonian as given in eq \ref{hamiltonian} depends on two coupling constants $\zeta$ and $\lambda$, where $\hat{V}$ is the intermonomer interaction operator, $\hat{F}$ is the unperturbed operator defined as the sum of monomer Fock operators, and $\hat{W}$ is the intramonomer correlation operator defined as the difference between the sum of monomer Hamiltonians and the sum of monomer Fock operators.

\begin{equation}\label{hamiltonian}\begin{aligned}\hat{H}(\zeta,\lambda) = \hat{F} + \zeta \hat{V} + \lambda \hat{W}\end{aligned}\end{equation}

The interaction energy is obtained by a twofold perturbation expansion given in eq \ref{pol_exch} where \textit{v} and \textit{w} define the orders of perturbation in $\zeta$ and $\lambda$.
Hence, at zeroth order, the SAPT solution consists of two non-interacting monomers, each of which is described by a separate Hartree-Fock (HF) wave function.

\begin{equation}\label{pol_exch}\begin{aligned}\Delta E^{\mathrm{SAPT}} = \sum_{v=1}^{\infty} \sum_{w=0}^{\infty} {\Delta E^{(vw)}}\end{aligned}\end{equation}

In our interaction energy decomposition, we apply the SAPT0\cite{sapt0_ref1,sapt0_ref2} method that contains only zeroth-order terms in $\lambda$ as given in eq \ref{eq:sapt0}. The jun-cc-pVDZ basis set\cite{seasonal} was used and all SAPT computations were performed with the Psi4 quantum chemistry software.\cite{psi4} The accuracy of SAPT0 as a qualitative computational method is also verified.

\begin{equation}\label{eq:sapt0}\begin{aligned}\Delta E^{\mathrm{SAPT0}} = \Delta E^{\mathrm{SAPT0}}_{\mathrm{elst}} + \Delta E^{\mathrm{SAPT0}}_{\mathrm{exch}} + \Delta E^{\mathrm{SAPT0}}_{\mathrm{ind}} + \Delta E^{\mathrm{SAPT0}}_{\mathrm{disp}} \end{aligned}\end{equation}

We introduce the field-dependent one-particle charge density of a monomer $M$ $\rho^M_{\mathrm{tot}} (\mathbf{r},\mathbf{r}',\boldsymbol{F})$ as given in eq \ref{dm_tot} with $\rho^M$ the monomer electron density and the fixed nuclear charge density expressed in terms of the nuclear charge $Z_I$ of atom \textit{I} and $\delta(\mathbf{r})$ the three-dimensional Dirac delta function.

\begin{equation}
\label{dm_tot}
\begin{aligned}
    \rho^M_{\mathrm{tot}} (\mathbf{r}_{1},\mathbf{r}_{1}',\boldsymbol{F})=
    -\rho^M (\mathbf{r}_{1},\mathbf{r}_{1}',\boldsymbol{F}) +\sum_{I \in M} Z_{I} \delta(\mathbf{r}_{1}'-\mathbf{r}_{I})
\end{aligned}
\end{equation}

The first-order (in $\zeta$) contribution to the field-dependent SAPT0 interaction energy $\Delta E^{(10)}$ contains the mean field electrostatic and exchange interaction of the two unperturbed monomer one-particle charge densities.
In the closed shell case, we can express this in a concise form as given in eq \ref{inten_first}, introducing the Coulomb interaction operator $\hat{\Tilde{V}}_{12} = |{\mathbf{r}_{1}-\mathbf{r}_{2}}|^{-1}\big(1-\frac12 \hat{P}'_{12}\big)$ for the 1-2 particle index pair with the $\hat{P}'_{12}$ permutation operator that interchanges the two electron indices, if both present, but evaluates to zero otherwise.

\begin{equation}
\label{inten_first}
\begin{aligned}
      & \Delta E^{(10)} (\boldsymbol{F})= \Delta E^{\mathrm{SAPT0}}_{\mathrm{elst}}(\boldsymbol{F}) + \Delta E^{\mathrm{SAPT0}}_{\mathrm{exch}}(\boldsymbol{F}) = \\
      & \int_{\mathbf{r}'_{1}=\mathbf{r}_{1}} \int_{\mathbf{r}'_{2}=\mathbf{r}_{2}} \hat{\Tilde{V}}_{12}  
      \rho^A_{\mathrm{tot}}(\mathbf{r}_{1},\mathbf{r}_{1}',\boldsymbol{F})
      \rho^B_{\mathrm{tot}}(\mathbf{r}_{2},\mathbf{r}_{2}',\boldsymbol{F})
      \mathrm{d^3}\mathbf{r}_{1}
      \mathrm{d^3}\mathbf{r}_{2}
\end{aligned}
\end{equation}

Note that the operator $\hat{\Tilde{V}}_{12}$ does not act on functions of primed variables, and the prime is removed only before integration. The exchange interactions arise via the contributions of $\hat{P}'_{12}$.
The one-particle electron density can be conveniently represented as matrix $(D_{rs})$ in the atomic orbital basis set.
Eq \ref{dens_mat} introduces the the monomer electron density matrix $(D_{rs})$. 

\begin{equation}
   \label{dens_mat}
   \rho^M (\mathbf{r},\mathbf{r}',\boldsymbol{F}) = \sum_{rs} D^{M}_{r s}(\boldsymbol{F})\phi^*_r(\mathbf{r}') \phi_s(\mathbf{r})
\end{equation}

The second and higher order contributions (in $\zeta$) to the SAPT0 interaction energy contain terms from the polarization of $\rho^M(\mathbf{r},\mathbf{r}')$ due to the intermolecular interaction.
Pure induction contributions $E^{\mathrm{SAPT0}}_{\mathrm{ind}}$ (i.\ e.\ pure static polarization interaction) up to the formally infinite order in $\zeta$ can be computed as the remaining HF contributions to the interaction energy as given in eq \ref{sapt0_ind}.

\begin{equation}
\label{sapt0_ind}
\begin{aligned}
    E^{\mathrm{SAPT0}}_{\mathrm{ind}} = \Delta E^{\mathrm{HF}} - \Delta E^{(10)} \quad .
\end{aligned}
\end{equation}

The lowest order dispersion contributions (second order in $\zeta$), arise from two-body dynamic polarization between the monomers' charge densities and are also taken into account in SAPT0. 
They contain a Coulomb interaction and an exchange interaction part and are given in eq \ref{sapt0_disp}.

\begin{equation}
\label{sapt0_disp}
\begin{aligned}
\Delta E^{\mathrm{SAPT0}}_{\mathrm{disp}} = \Delta E^{(20)}_{\mathrm{disp}} + \Delta E^{(20)}_{\mathrm{exch-disp}} \quad .
\end{aligned}
\end{equation}

The components of the SAPT0 interaction dipole moment $\Delta \mu^{(10)}$ and interaction polarizability $\Delta\alpha_{ij}^{(10)}$ can be expressed in terms of the monomer charge densities $\rho^M_{\mathrm{tot}}(\mathbf{r},\mathbf{r}',\boldsymbol{F})$ by inserting eq \ref{inten_first} into  eqs \ref{intmu} and \ref{intalpha}.
Since the Coulomb interaction operator $\hat{\Tilde{V}}_{12}$ does not depend on the external field, the field dependence of $\Delta E^{(10)}$ can be expressed in the set of atomic orbital density matrix elements $\{D^{A}_{r s}(\boldsymbol{F}), D^{B}_{r s}(\boldsymbol{F})\}$, given for the electron-electron interaction part of $\Delta E^{(10)}$ in eq \ref{eq:E_AO}.

\begin{equation}
   \label{eq:E_AO}
   \begin{aligned}
   \Delta & E^{(10)}_{\mathrm{el-el}} (\boldsymbol{F})= \sum_{rstu} D^{A}_{r s} (\boldsymbol{F}) \ D^{B}_{t u} (\boldsymbol{F}) \\
   & \int_{\mathbf{r}'_{1}=\mathbf{r}_{1}} \int_{\mathbf{r}'_{2}=\mathbf{r}_{2}} \phi^*_r(\mathbf{r}_1') \phi_s(\mathbf{r}_1) \hat{\Tilde{V}}_{12} \phi^*_t(\mathbf{r}_2') \phi_u(\mathbf{r}_2) \mathrm{d}^3\mathbf{r}_1 \mathrm{d}^3\mathbf{r}_2
   \end{aligned}
\end{equation}

Using the chain rule and introducing the short-hand notations $\partial_{(Mrs)} \Delta E^{(10)} \equiv \frac{\partial \Delta E^{(10)}(\boldsymbol{F})}{\partial D^M_{rs}(\boldsymbol{F})}\bigr|_{\boldsymbol{F}=\boldsymbol{0}}$,
$\partial_{(Mrs)(Ntu)} \Delta E^{(10)} \equiv \frac{\partial^2 \Delta E^{(10)}(\boldsymbol{F})}{\partial D^M_{rs}(\boldsymbol{F})\partial D^N_{tu}(\boldsymbol{F})}\bigr|_{\boldsymbol{F}=\boldsymbol{0}}$, 
$\partial_i D^M_{rs} \equiv \frac{\partial D^M_{rs}(\boldsymbol{F})}{\partial F_i}\bigr|_{\boldsymbol{F}=\boldsymbol{0}}$, and $\partial_{ij} D^M_{rs} \equiv \frac{\partial^2 D^M_{rs}(\boldsymbol{F})}{\partial F_i \partial F_i}\bigr|_{\boldsymbol{F}=\boldsymbol{0}}$ 
one can express $\Delta \mu^{(10)}_{i} $ and $\Delta \alpha^{(10)}_{ij}$ in terms of monomer-resolved density polarizations in the electric field as given in eqs \ref{intmu_mono} and \ref{intalpha_mono}.

\begin{equation}\label{intmu_mono}\begin{aligned}
\Delta \mu^{(10)}_{i}= - \sum_{rs} \big[  \partial_{(Ars)} \Delta E^{(10)} \partial_i D^A_{rs} + 
 \partial_{(Brs)} \Delta E^{(10)} \partial_i D^B_{rs} \big]
\end{aligned}\end{equation}

\begin{equation}\label{intalpha_mono}\begin{aligned}
\Delta \alpha^{(10)}_{ij}= - \sum_{rs} \big[ & \partial_{(Ars)} \Delta E^{(10)}  \partial_{ij} D^A_{rs} + 
 \partial_{(Brs)} \Delta E^{(10)}  \partial_{ij} D^B_{rs} \big] \\
- \sum_{rstu}  \Big[ & \partial_{(Ars)(Btu)} \Delta E^{(10)} 
  \big[ \partial_i D^A_{rs} \partial_j D^B_{tu} + 
 \partial_j D^A_{rs} \partial_i D^B_{tu} \big] \ \Big]
\end{aligned}\end{equation}

Hence, there are two separate contributions to $\Delta \mu^{(10)}_{i}$, arising from the linear polarization of monomer A and B, respectively.
There are contributions to $\Delta \alpha^{(10)}_{ij}$ originating from quadratic polarization on only one monomer, as well as from ``mixed'' contributions that stem from the combined linear polarization of monomers A and B.
Electrostatic and exchange monomer contributions to the SAPT0 interaction-induced electric properties can be obtained separately (cf.\ eq \ref{inten_first}).
The monomer decomposition analysis of the first-order terms enables a better qualitative understanding of the link between monomer electric properties and interaction induced electric properties.
It is in this context interesting to note that there is a ``twofold'' physical interpretation of the interaction induced electric properties,\cite{heijmen1996} e.\ g.\ for the interaction dipole moment: First, one can see it as an expectation value of the dimer due to the mutual charge polarization among the monomers; second, one can see it as the first order change in the interaction energy due to the monomer charge density polarization by the external field, as expressed in eq \ref{intmu_mono}.

\section{Result and discussion}
\subsection{CCSD(T) interaction-induced properties along the \textit{z}-axis}

\begin{figure*}[t!]\centering\includegraphics[width=1.0\linewidth]{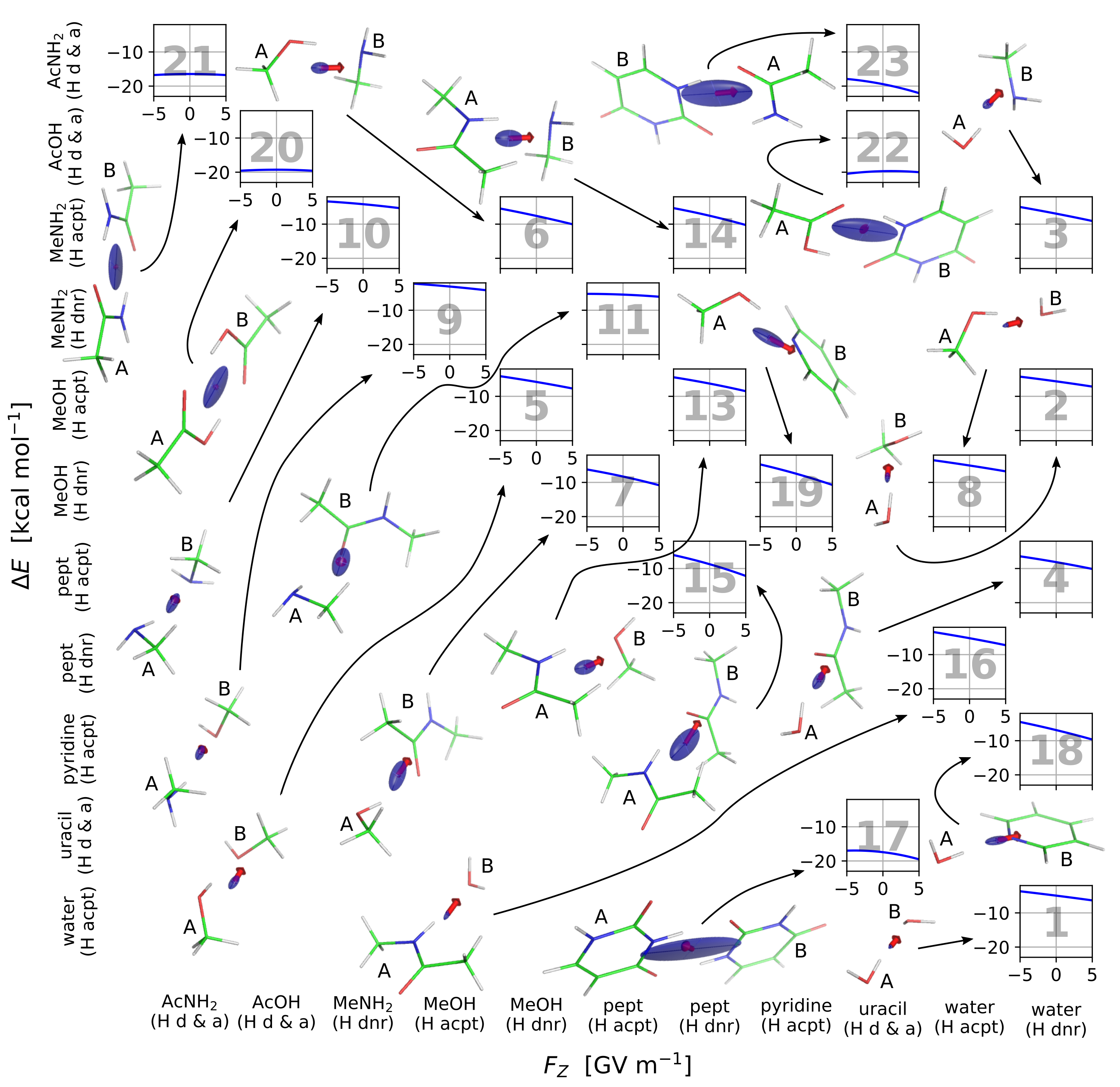}
\caption{\label{selected1} 
Interaction energy of hydrogen bonded dimers in the S66 set as a function of $F_z$ with the dimer number given in the plot background. 
The plots are arranged in a matrix with rows and columns expanding the monomer conformations in the respective combination of a dimer. 
Brackets denote ``H dnr'' for H-bond donor, ``H acpt'' for H-bond acceptor, ``H d \& a'' for H-bond donor and acceptor. 
Dimer 12 not shown since it would overlap with dimer 3 within our matrix arrangement. 
Corresponding 3D structural depictions, linked through black arrows, visualize the interaction dipole moment vector (red arrow) and the interaction polarizability principal components (blue ellipsoid with black and white rods for positive and negative eigenvalues, respectively) computed at the dRPA@PBE0/aug-cc-pVQZ level of theory. 
The 3D structures were created using the PyMol software package.\cite{PyMOL}}
\end{figure*}

\begin{figure*}[t!]\centering\includegraphics[width=1.0\linewidth]{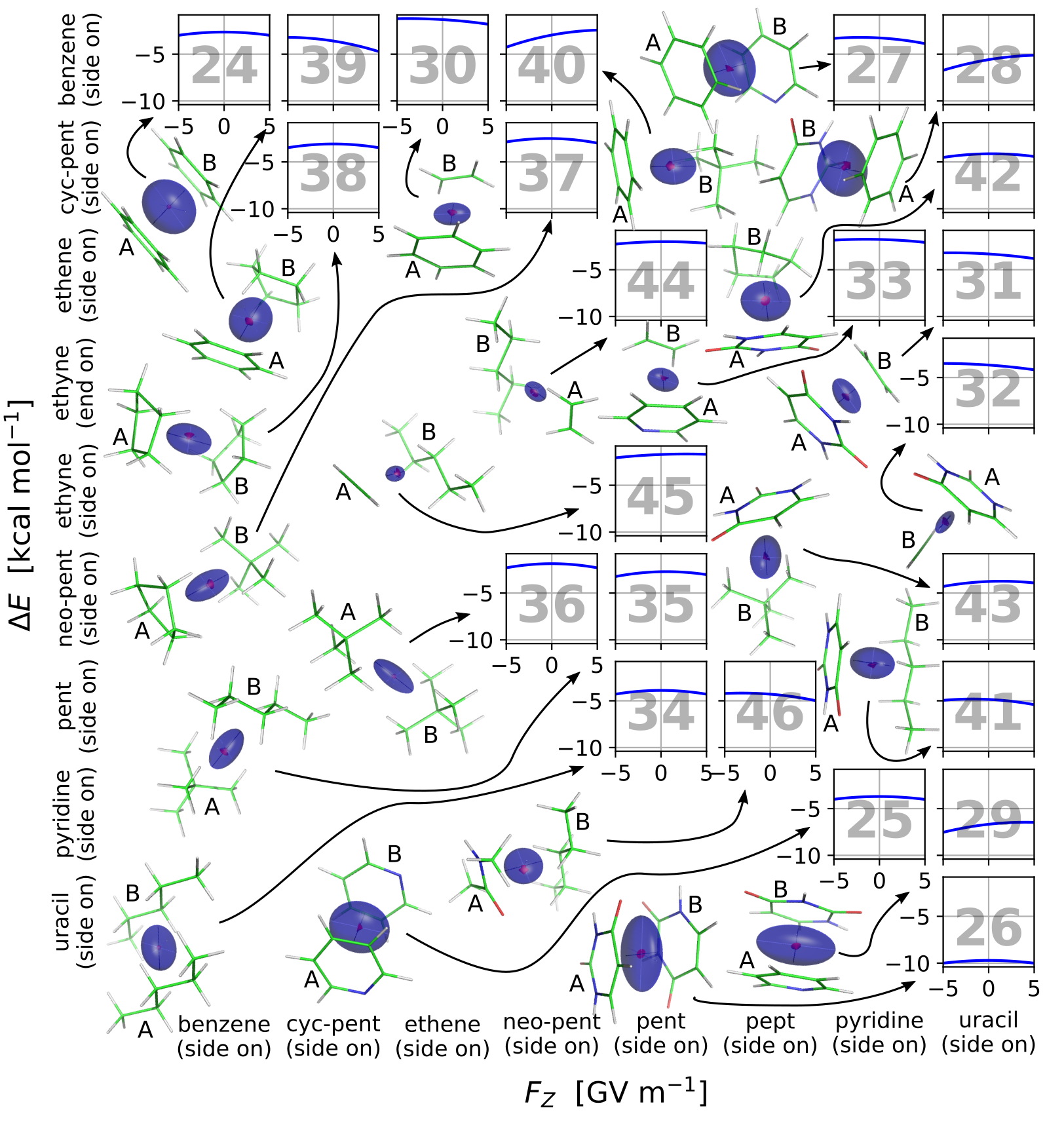}\caption{\label{selected2} 
Interaction energy of $\pi$-stacking and London dispersion bonded dimers in the S66 set as a function of $F_z$ with the dimer number given in the plot background. 
The plots are arranged in a matrix with rows and columns expanding the monomer conformations in the respective combination of a dimer. 
Brackets denote ``side on'' and ``end on'' for parallel and orthogonal relative placements, respectively. 
Corresponding 3D structural depictions, linked through black arrows, visualize the interaction dipole moment vector (red arrow) and the interaction polarizability principal components (blue ellipsoid with black and white rods for positive and negative eigenvalues, respectively) computed at the dRPA@PBE0/aug-cc-pVQZ level of theory. 
}\end{figure*}

\begin{figure*}[t!]\centering\includegraphics[width=0.9\linewidth]{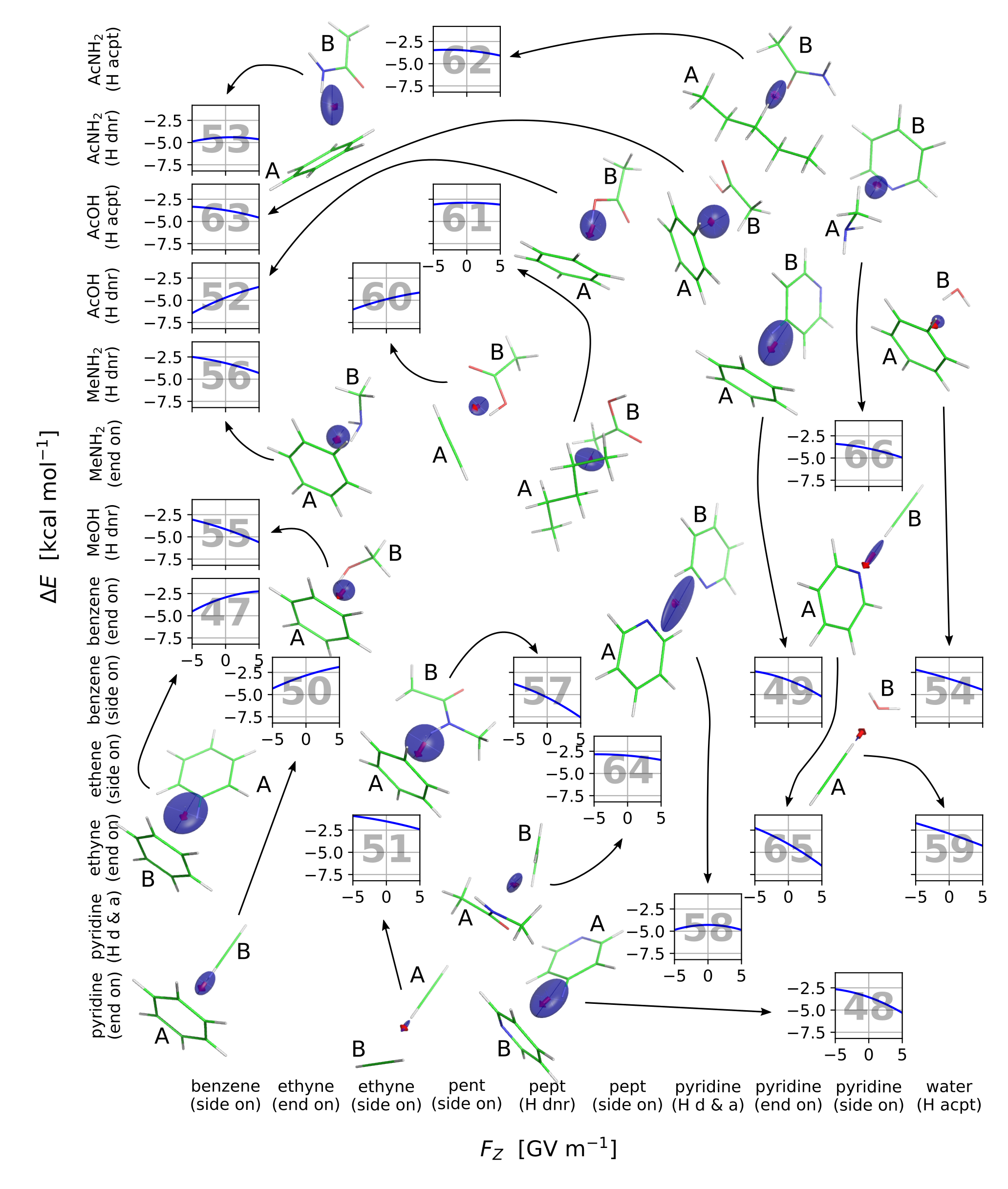}
\caption{\label{selected3} Interaction energy of mixed interaction bonded dimers in the S66 set as a function of $F_z$ with the dimer number given in the plot background. 
The plots are arranged in a matrix with rows and columns expanding the monomer conformations in the respective combination of a dimer. 
Brackets denote ``H dnr'' for H-bond donor, ``H acpt'' for H-bond acceptor, ``H d \& a'' for H-bond donor and acceptor.
Corresponding 3D structural depictions, linked through black arrows, visualize the interaction dipole moment vector (red arrow) and the interaction polarizability principal components (blue ellipsoid with black and white rods for positive and negative eigenvalues, respectively) computed at the dRPA@PBE0/aug-cc-pVQZ level of theory.} 
\end{figure*}

The interaction energies of the S66 set of non-covalent dimers are computed as a function of the OEEF categorized as "H-bonding" (dimers 1-23), "$\pi$-stacking" (dimers 24-33), "London dispersion" (dimers 34-46), and "mixed interactions" (dimers 47-66).
All dimers are depicted as 3D structures in Figure \ref{selected1} (hydrogen bonded dimers), Figure \ref{selected2} ($\pi$-stacking and London dispersion), and Figure \ref{selected3} (mixed interaction).
The dimers are in these Figures arranged in matrices consisting of monomer interaction types in its rows and columns, so that the relation between the dimers in terms of monomer composition and conformation is visualized.

According to our CCSD(T) results, the magnitude of the first interaction hyperpolarizability along the $z$-axis is usually small (0.165 \AA$^5$ on average obtained by a polynomial fit up to third order which means a 0.002 kcal mol$^{-1}$ effect in the interaction energy at our largest field strength).
Hence, the interaction energy as a function of the OEEF is fully determined by the zero field interaction energy, interaction dipole moment, and interaction polarizability.
For further details on the numerical fitting procedure we refer to the SI.

\begin{figure}[h!]\centering\includegraphics[width=1.09\linewidth]{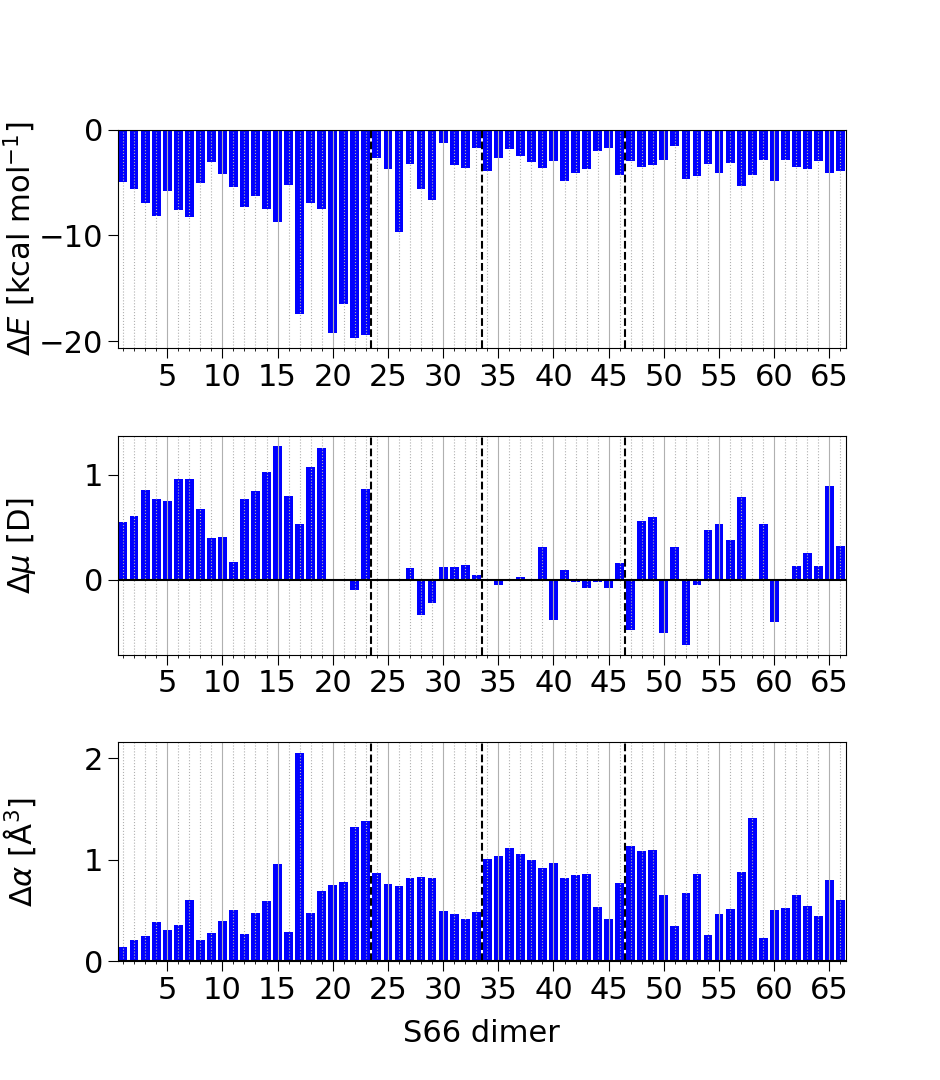}\caption{\label{ccsdt}CCSD(T)-F12b/aug-cc-pVDZ interaction energies, dipole moments, and polarizabilities for the S66 noncovalent dimers along the \textit{z}-axis. Vertical black dashed lines separate hydrogen bonding, $\pi$-stacking, London dispersion, and mixed interaction classes.}
\end{figure}

The reference CCSD(T) interaction energies, dipole moments, and polarizabilities along the $z$-axis are shown in Figure \ref{ccsdt}.
Among the interaction types, the magnitude of the interaction energy is large in the case of hydrogen bonding ($\approx$9 kcal mol$^{-1}$ on average), while moderately large in the other cases ($\approx$3-4 kcal mol$^{-1}$ on average).

We observe positive interaction dipole moments in most of the cases, which means the interaction-induced polarization enhances the monomer polarity.
However, negative interaction dipole moments occur for the hydrogen bonded AcOH - uracil dimer, and for some of the dimers corresponding to the London dispersion and mixed interaction classes.
Furthermore, (near) zero interaction dipole moments occur for some homodimers due to symmetry (or near-symmetry).
The largest magnitudes of interaction dipole moments occur for hydrogen bonded systems ($\approx$0.7 D on average), dispersion bonded cases show rather small magnitudes ($\approx$0.1 D on average), and mostly moderate magnitudes are found in the mixed influenced cases ($\approx$0.4 D on average).
The large interaction dipole moment for asymmetric hydrogen bonded systems (e.\ g.\ dimers 1, 15, and 17) or hydrogen-$\pi$-interacting systems (e.\ g.\ dimers 47 and 57) suggests that the OEEF can considerably affect the interaction energy in ice, proteins, or double stranded DNA.\\
\indent Parallel displaced and T-shaped $\pi$-stacking homodimers (e.\ g.\ dimers 24 and 47) have similar interaction energies. The former however have a near zero interaction dipole moment along the $z$-axis while the latter can bear a significant interaction dipole moment.
This may be utilized for selective growth of molecular crystals with fishbone lattice structures from melts in strong OEEFs.\\
\indent In most of the cases, the magnitude of the interaction polarizability is small ($\approx$0.6-0.7 \AA$^3$ on average), and it has only a small effect on the shape of the interaction energy function with respect to the field strength. Its influence is somewhat more important in the $\pi$-$\pi$-interaction and London dispersion dominated dimers ($\approx$0.9 \AA$^3$ on average), where the interaction dipole moment is small.
The largest interaction polarizabilities occur in dimers of aromatic heterocycles with the ring planes oriented along the $z$ axis (dimers 17 and 58) with an increase in interaction energy of up to 0.87 kcal mol$^{-1}$ at the largest considered field strength.

\subsection{SAPT0 decomposition of the interaction-induced properties along the \textit{z}-axis}

\begin{figure}[h!]\centering\includegraphics[width=1.09\linewidth]{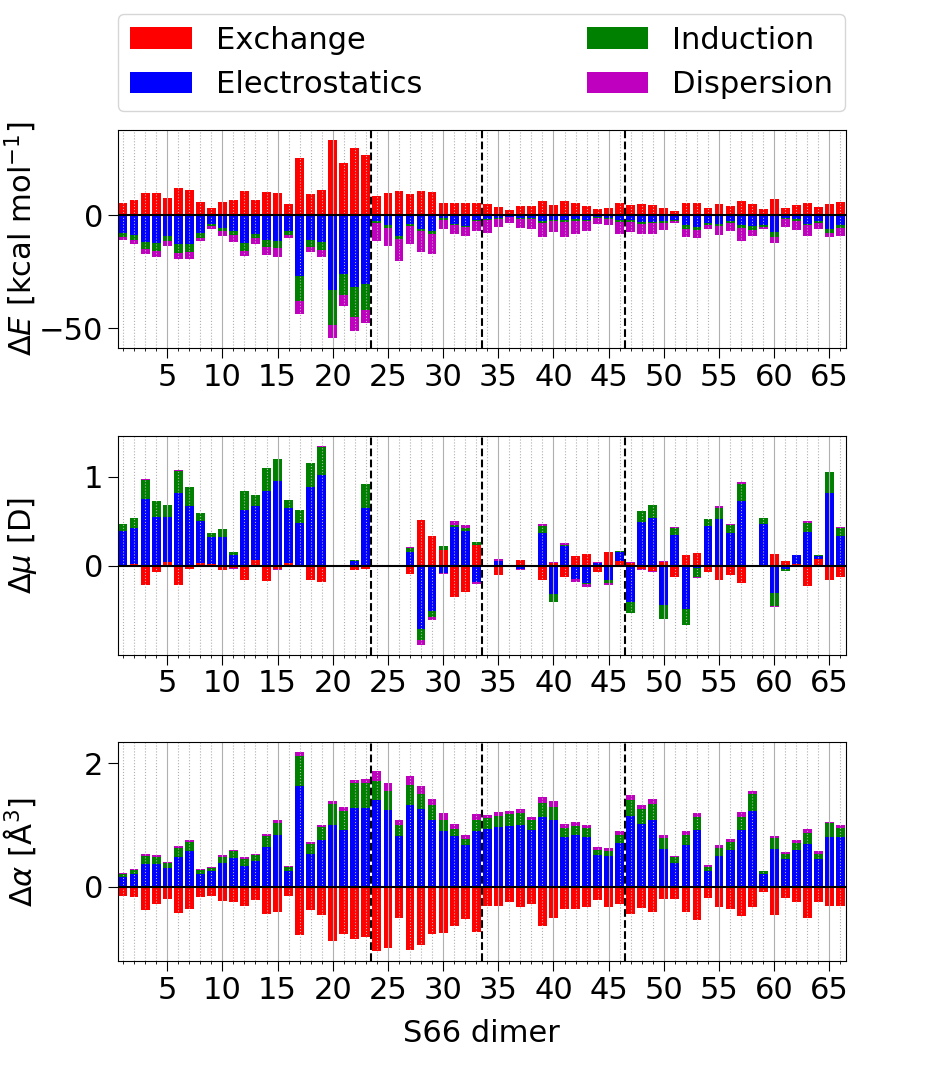}\caption{\label{fig:sapt0}SAPT0 decomposition of the interaction energy, dipole moment, and polarizability for the S66 noncovalent dimers along the \textit{z}-axis. Vertical black dashes lines separate hydrogen bonding, $\pi$-stacking, London dispersion, and mixed interaction classes.}\end{figure}

For a qualitative understanding of the most relevant physical interactions, we compute the SAPT0 decomposition of the interaction induced electric properties. 
Interaction induced properties of Watson-Crick amino acid pairs have also been studied with higher orders of SAPT\cite{czyznikowska2013} and also with taking into account nuclear relaxation effects for small hydrogen- and halogen-bonded dimers in the order up to interaction first hyperpolarizabilities.\cite{medved2020,zalensky2018} 
The SAPT0 mean absolute errors for the interaction energy, dipole moment, and polarizability are, respectively, 0.50 kcal mol$^{-1}$, 0.039 D, and 0.15 {\AA}$^3$.
Hence, SAPT0 is in qualitative agreement with the coupled cluster reference.\\
\indent Figure \ref{fig:sapt0} depicts the SAPT0 electrostatic, exchange, induction, and dispersion contributions for $\Delta E$, $\Delta \mu_z$, and $\Delta \alpha_z$.
In the case of hydrogen bonding, all four interaction contributions to the interaction energy are significant.
The sum of the (first-order) electrostatic and exchange parts is usually attractive; however, both the induction and dispersion parts are larger in magnitude than the first-order interaction energy.
In case of dispersion bonded dimers the first-order interaction energy is repulsive and the dimers are bound by the dominating attractive dispersion interaction.\\
\indent Interestingly, the contributions to the interaction dipole moment strongly differ from those to the interaction energy.
In the hydrogen bonded dimers, the interaction dipole moment is always positive and dominated by the electrostatic part.
Exchange contributions are rather small and tend to slightly moderate the electrostatic contribution. Higher-order static polarization effects described by the induction contributions further enhance the interaction dipole moment.
Intermolecular dispersion at the SAPT0 level of theory gives very small contributions.
In the $\pi$-stacking and London dispersion bonded dimers, the electrostatic part of the interaction dipole moment is quenched by the exchange part to some extent, and the direction of the interaction dipole moment does not seem to correlate with the direction of the sum of the monomer dipole moments. The induction and intermolecular dispersion contributions are small for SAPT0.
It should be pointed out here that SAPT0 does not cover higher-order terms that include many-body and intramonomer dispersion.
For the interaction polarizability of cytosine dimers in stacked alignments, it was indeed found that electrostatic effects from intramonomer correlation are even larger than the lowest order intermolecular dispersion contributions.\cite{czyznikowska2013} \\
\indent Regardless the interaction class, the interaction polarizability contains important static polarization contributions in the lowest order (electrostatics and exchange) as well as at higher orders (induction).
The SAPT0 dispersion contribution is found to be only significant in the case of $\pi$-$\pi$-interactions. 
Czy{\.z}nikowska \emph{et al.}\cite{czyznikowska2013} found for Watson-Crick amino acid pairs very strong exchange and induction contributions to the interaction polarizability which underlines the exceptionally strong interaction in these dimer systems.  
Our findings are in agreement with an earlier computation for the HCN dimer where the interaction polarizability was found to be greatly influenced by electrostatic effects.\cite{dft_interaction_induced3}

\subsection{SAPT0 monomer contributions to the first-order interaction-induced properties along the \textit{z}-axis}

\begin{figure}[h!]\centering\includegraphics[width=1.1\linewidth]{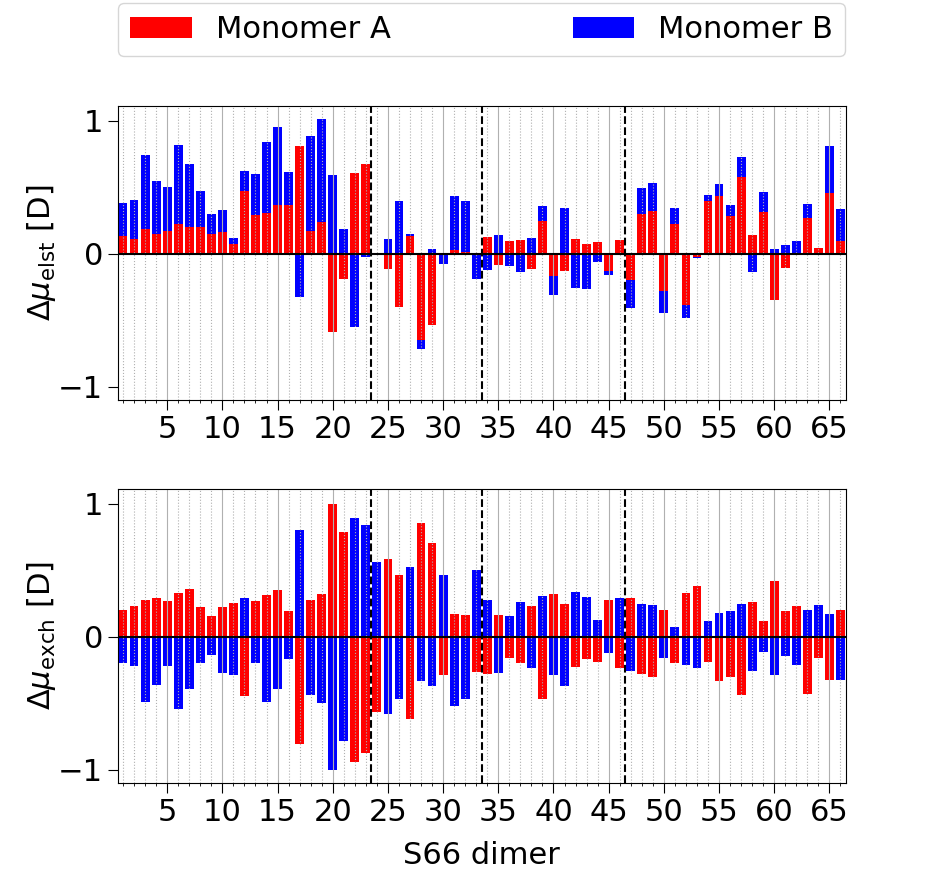}\caption{\label{mono_dip}Monomer first-order polarization contributions to the SAPT0 electrostatic (top panel) and exchange interaction (bottom panel) dipole moments for the S66 noncovalent dimers along the \textit{z}-axis. Vertical black dashed lines separate hydrogen bonding, $\pi$-stacking, London dispersion, and mixed interaction classes.}\end{figure}

\begin{figure}[h!]\centering\includegraphics[width=1.1\linewidth]{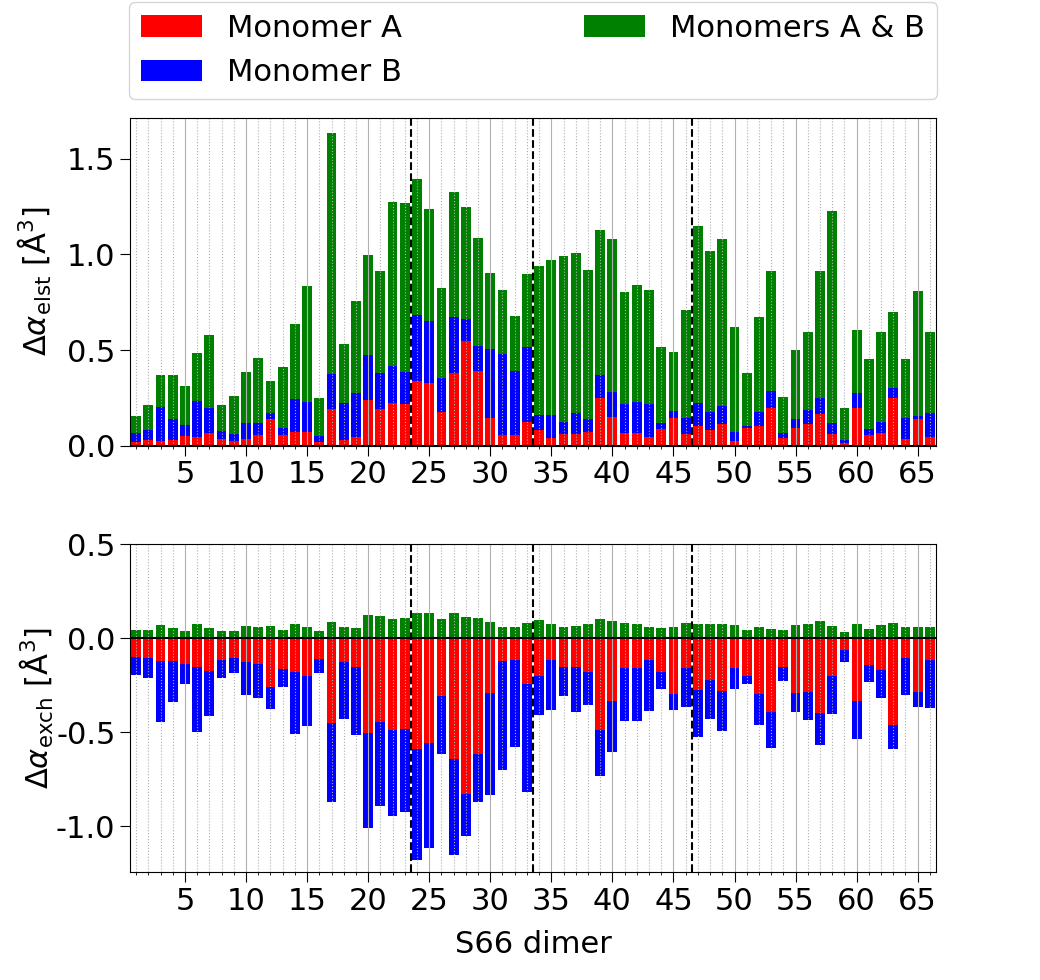}\caption{\label{mono_pol}Monomer second-order polarization and combined monomer first-order polarization contributions to the SAPT0 electrostatic and exchange interaction polarizabilities for the S66 noncovalent dimers along the \textit{z}-axis. Vertical black dashed lines separate hydrogen bonding, $\pi$-stacking, London dispersion, and mixed interaction classes.}\end{figure}

A link between monomer electric properties and interaction-induced electric properties is provided for the SAPT0 electrostatic and exchange energy contributions by eqs \ref{intmu_mono} and \ref{intalpha_mono}.
The monomers are assigned the labels ``A'' and ``B'' as indicated in the 3D structures of the dimers in Figure \ref{selected1} (hydrogen bonded dimers), Figure \ref{selected2} ($\pi$-stacking and London dispersion), and Figure \ref{selected3} (mixed interaction).
Figures \ref{mono_dip} and \ref{mono_pol} show the monomer contributions to the interaction dipole moment and the interaction polarizability, respectively.
In singly hydrogen bonded dimers (dimers 1-16 and 18-19), the hydrogen bond donor (monomer A) and acceptor (monomer B) contributions to the electrostatic interaction dipole moment have the same sign.
The donor contribution to the electrostatic interaction dipole moment is relatively less pronounced in the case of OH groups (31\% on average with the water and methanol H-bond donors), while somewhat more pronounced in the case of NH groups (53\% on average with the peptide and methyl amine H-bond donors).
In (nearly or fully) symmetric doubly hydrogen bonded dimers (dimers 17 and 20--23) or $\pi$-stacking/London dispersion dominated homodimers (dimers 24-26, 34, 36, and 38), the oppositely signed monomer contributions to the electrostatic interaction dipole moment
mostly cancel each other.
The directionality of hydrogen bonds is thus a strong vector for designing the first-order interaction electric response.
The monomer contributions to the exchange interaction dipole moment always have the opposite sign and approximately (or exactly, in case of symmetric homodimers) the same magnitude.
This is a manifestation of the increase in Pauli repulsion when the monomer electron density polarizes towards the other monomer. 
As this results generally in only a small net exchange contribution to the interaction dipole moment, Pauli repulsion appears to act in a rather ``symmetric'' way upon monomer polarization.
This observation holds for all dimer systems with the exception of some $\pi$-stacking dimers (dimers 28--33).
It is apparent that the interaction-induced static polarization of the hydrogen bond donor and acceptor is the main contribution to large interaction dipole moments in the hydrogen-bonded cases.
The hydrogen donor-acceptor conformation in these dimers leads to a strong electrostatic contribution to the interaction energy (cf.\ Figure \ref{fig:sapt0}) and rather short dimer distances where the charge distributions of the monomers arrange optimally one to each other.
This plays in favour of a strong monomer electrostatic polarization.
For the systems with $\pi$-$\pi$ interactions and London dispersion interaction, the picture is more heterogeneous regarding the electrostatic monomer contributions; they may be of the same or opposite sign and are quite large in magnitude for some cases. \\
\indent The electrostatic interaction polarizability is composed only of positive contributions for both monomer second-order polarization and combined monomer first-order polarization.
For the monomer second-order polarizations this is an interesting finding, as it suggests that the monomer static polarizability always increases along the axis of interaction, i.\ e.\ the monomer polarization due to the interaction always leads to a more polarizable monomer electron density.
These monomer second-order polarizations are especially strong in hydrogen-bonded and $\pi$-stacking dimers.
It is remarkable that monomer first hyperpolarizabilites apparently play an important role, as they are at the origin of the monomer second order polarization, while interaction first hyperpolarizabilities are observed to be insignificant.
However, in the case of exceptionally ``field-sensitive'' systems, such as electron-rich, polar, strongly hydrogen-bonded dimers that are aligned along the $z$-axis, significant interaction first hyperpolarizabilities have been predicted.\cite{gora2013}\\
\indent In London dispersion and mixed interaction dimers the combined monomer polarization dominates mostly and is large for all dimers.
This is to be expected, as the ``cooperative'' polarization of both monomers under the electric field $F_z$ increases the charge density mobility across the dimer, as compared to the isolated monomer charge density mobility.
Remarkably, the monomer coupling term in the electrostatic interaction polarizability $\Delta\alpha_{zz}$ correlates well with the product of the corresponding monomer polarizabilities ($R^2 \approx 0.92$, see the SI for numerical details).
This proportionality may be a suitable simple design principle for the second-order behavior of the interaction energy in OEEFs in direction of the interaction axis $z$.\\
\indent The individual monomer contributions to the exchange interaction polarizability dominate by far and are negative.
This can be understood by considering that the repulsive intermolecular exchange interaction diminishes the polarization capacity of the individual monomer charge densities. 
The monomer coupling contributions to the exchange interaction polarizability give small positive contributions, indicating that slight cooperative effects in monomer polarization capacity may also be mediated by the exchange interactions.   

\subsection{Performance of density functional approximations for interaction induced electric properties}

The shortcomings (and corrections) of many of the original density functional approximations for van der Waals energies are well established. 
Their performance for interaction dipole moments and polarizabilities, however, is not that known.
While the accuracy of calculated electrostatic dipole (and multiple) moments of monomers is typically quite decent~\cite{ChemistsGuidetoDFT}, 
the predictive power of density functionals can vary dramatically when it comes to 
polarizability~\cite{van1996improved,champagne1998assessment,willow2016mp2,hait2018accurate,akter2021well}.
In order to facilitate future studies dealing with improved density functional approximations, we therefore have also included
a selection of common DFT approximations in this study, and we now discuss the corresponding numerical estimates 
of van der Waals interaction energies, dipole moments, and polarizabilities. 

\begin{figure}[h!]\centering\includegraphics[width=1.1\linewidth]{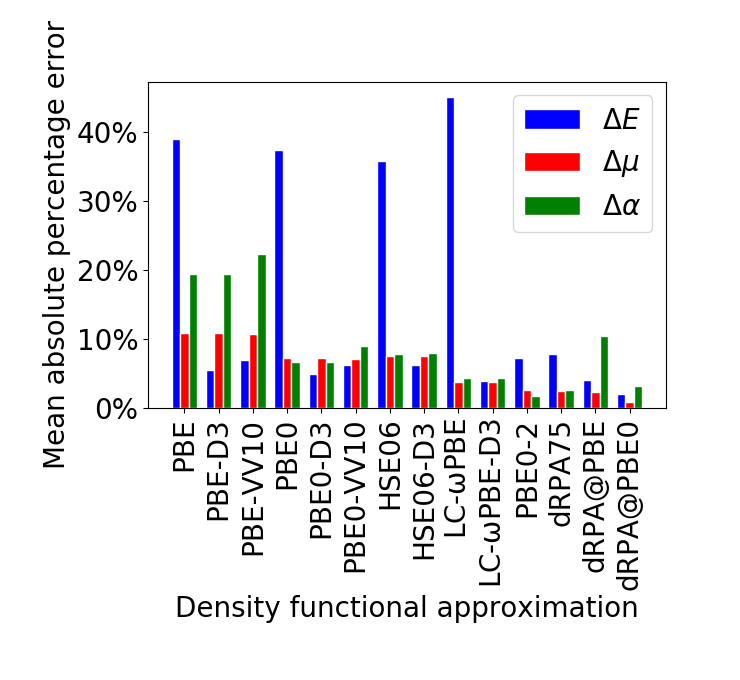}\caption{\label{dft} Mean absolute percentage error (MAPE) of the interaction energies, dipole moments, and polarizabilities obtained with various density functional approximations for the S66 noncovalent dimers along the \textit{z}-axis relative to the average magnitude of the same properties calculated with the CCSD(T)-F12b/aug-cc-pVDZ reference method.}\end{figure}

We have placed our main focus on results related to the PBE semilocal exchange-correlation kernel, augmented by various nonlocal exchange and correlation ingredients.
In terms of methodology aiming at the quantitative predictive power needed to account properly for long-range dispersion, 
we decided to focus our DFT benchmark study on this latter part. 
Figure \ref{dft} shows the mean absolute percentage error (MAPE) of the considered density functionals, including purely semilocal PBE and PBE0 for comparison.
Semilocal functionals like PBE cannot capture long-range dispersion interactions due the nonlocal nature of the latter.
The D3 dispersion correction can give a satisfactory description of long-range dispersion in some use cases, but is insensitive to the electric field. 
The VV10 dispersion correction is based on the electron density, but its effect on the interaction dipole moments is quite small, and the interaction polarizabilities can become even more erroneous if the calculated electron density is too diffuse.
The exact exchange mixing in hybrid functionals significantly lowers the interaction dipole moments and polarizabilities errors.
This may indicate that the quality of the calculated electron density is better compared to the electron densities obtained by purely semilocal functionals (the latter being too delocalized and diffuse due to the large self-interaction error).
However, an additional dispersion correction is clearly needed for reliable predictions.
The interaction-induced properties obtained with the screened hybrid HSE06 functional 
are very similar to the ones with the global hybrid PBE0 with the same fraction of exact exchange at short-range, due to the weak screening by the very small range-separation parameter.
The long-range correction in the LC-$\omega$PBE functional considerably improves the interaction dipole moments and polarizabilities, but such functionals still require a dispersion correction.
Our observations up to this point are in agreement with the literature, where the long-range corrected LC-BLYP method\cite{b88,lyp,lc_scheme} was reported to yield more accurate interaction-induced properties than the semilocal BLYP method\cite{b88,lyp} for hydrogen bonded dimers,\cite{dft_interaction_induced3} and global hybrid density functionals were found to be qualitatively correct for interaction dipole moments and mean interaction polarizabilities, but not for the anisotropy of the interaction polarizabilities for noble gas dimers.\cite{dft_interaction_induced4}
Our study suggests that long-range corrected functionals are generally more accurate than global hybrid ones for interaction-induced properties even though global hybrids can be occasionally better than long-range corrected hybrids depending on the parent semilocal functional.\cite{dft_interaction_induced2}
The double hybrid PBE0-2 functional with partial second-order perturbative correlation and a large fraction of exact exchange delivers somewhat less accurate interaction energies but more accurate interaction dipole moments and polarizabilities.
The dual-hybrid dRPA75 functional with full dRPA correlation and a similarly large fraction of exact exchange yields similarly accurate results.
Finally, the dRPA method with full exact exchange and dRPA correlation (dRPA@PBE0) leads to very accurate results for all three quantities as compared to the coupled cluster reference (MAPE$\le$3\% for all three quantities).
Nevertheless it should be noted that the dRPA interaction dipole moments and polarizabilities are quite sensitive to the reference orbitals: Omitting the exact exchange (dRPA@PBE) introduces considerable errors to the interaction polarizabilities. The sensitivity of density functional approximations with respect to orbital relaxation and higher-order long-range dispersion effects was also found for the computation of electric properties of endohedral small polar molecules in carbon nanotubes.\cite{kozlowska}

In summary, the LC-$\omega$PBE functional is a surprisingly reliable method given its low computational cost (MAPE$\le$5\% for all three quantities), but other computationally efficient methods (PBE0-D3, PBE0-VV10, HSE06-D3) also perform fairly well (MAPE$\le$10\% for all three quantities).

\subsection{Interaction-induced properties in three dimensions}

Given the good performance of dRPA@PBE0/aug-cc-pVQZ with respect to the coupled cluster reference, this level of theory was chosen to compute all the elements of the interaction dipole moment vector and interaction polarizability matrix with a three-dimensional grid (as described in the methods section) that allows for fitting of a quadratic function.
The interaction dipole moments are depicted in vector form (direction and magnitude) for the S66 dimers in Figure \ref{selected1} (hydrogen bonded dimers), Figure \ref{selected2} ($\pi$-stacking and London dispersion), and Figure \ref{selected3} (mixed interaction).
The interaction polarizabilities with directions and magnitude of the three principal components depicted as ellipsoids are given in the same Figures.
We tested the quadratic dependency of the interaction energy with respect to the OEEF for the CCSD(T) reference method also in the \textit{xy}-plane on a denser grid for some selected dimers.
The magnitude of the first interaction hyperpolarizability components is in this case 0.046 \AA$^5$ on average, obtained by a polynomial fit up to third order and resulting in a 0.0007 kcal mol$^{-1}$ effect on the interaction energy at our largest field strength (further details are given in the SI).
Hence, the first interaction hyperpolarizability appears also to be negligible in the $xy$-plane.

\begin{figure}[h!]\centering\includegraphics[width=1.11\linewidth]{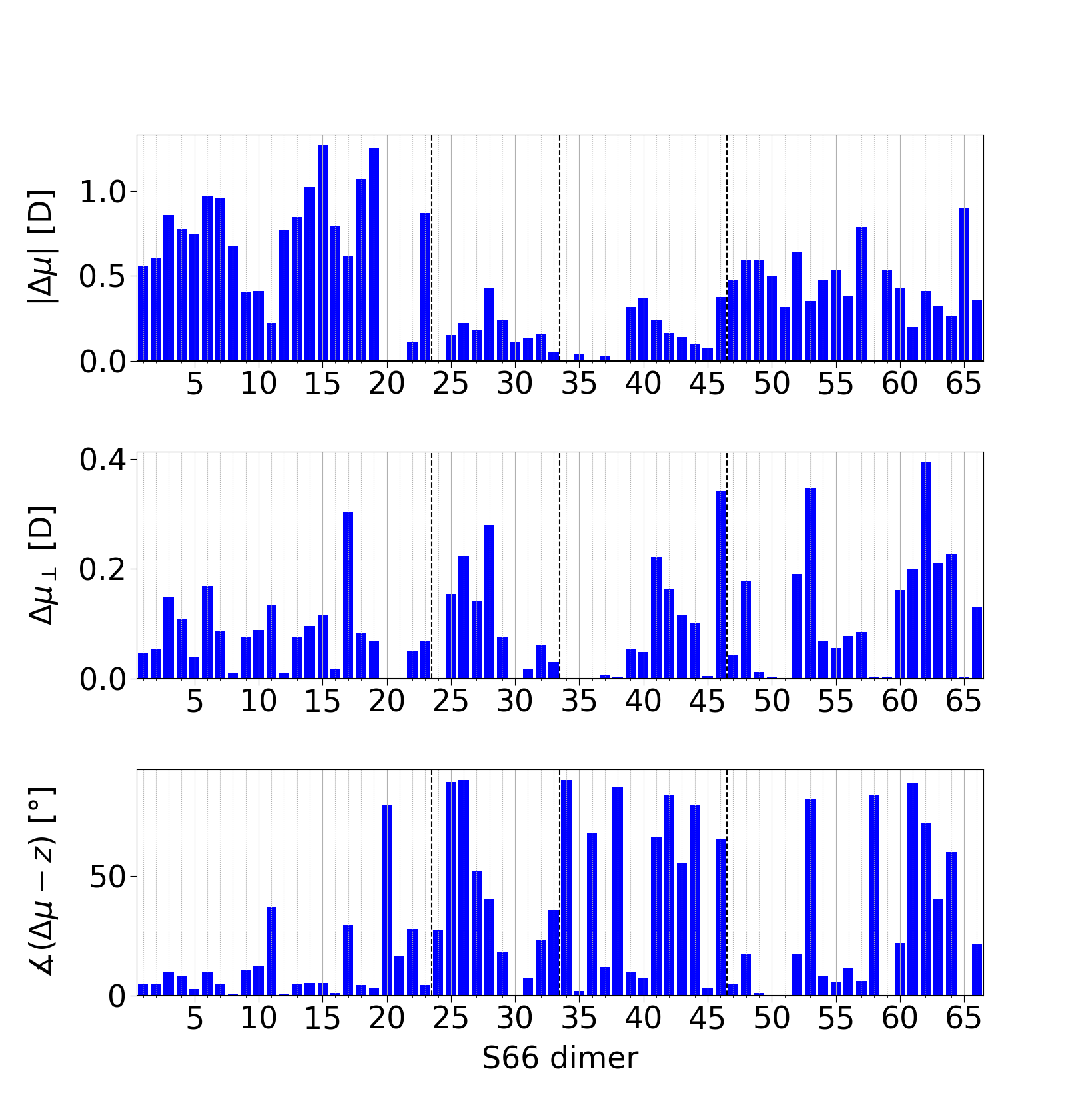}\caption{\label{dip}
Interaction dipole moments for the S66 noncovalent dimers calculated with dRPA. 
Magnitude of the interaction dipole moment vector (top), 
interaction dipole moment component perpendicular to the \textit{z}-axis  (mid), and
angle of the interaction dipole moment vectors with respect to the \textit{z}-axis (bottom). 
Vertical black dashed lines separate hydrogen bonding, $\pi$-stacking, London dispersion, and mixed interaction classes.}\end{figure}

The average value of the angle of the interaction dipole moment vector with respect to the \textit{z}-axis is 28.0\degree{} for the whole database, while it is 12.6\degree{} for hydrogen bonding, 38.4\degree{} for $\pi$-$\pi$-interaction, and 48.4\degree{} for London dispersion dominated dimers.
The individual angles for the S66 dimers are given in Figure \ref{dip}.
The dominant interaction dipole moment component along the $z$-axis for H-bonded systems is a result of the hydrogen bond's directionality.
In the case of $\pi$-$\pi$- and London dispersion interactions the orientation of the interaction dipole moment mainly originates from the monomers' orientation, as the lowest-order contribution to the $\Delta\mu_i$ vector component depends on the monomer polarizations $\partial_i D^M_{rs}$ (see eq \ref{intmu_mono}).

\begin{figure}[h!]\centering\includegraphics[width=1.11\linewidth]{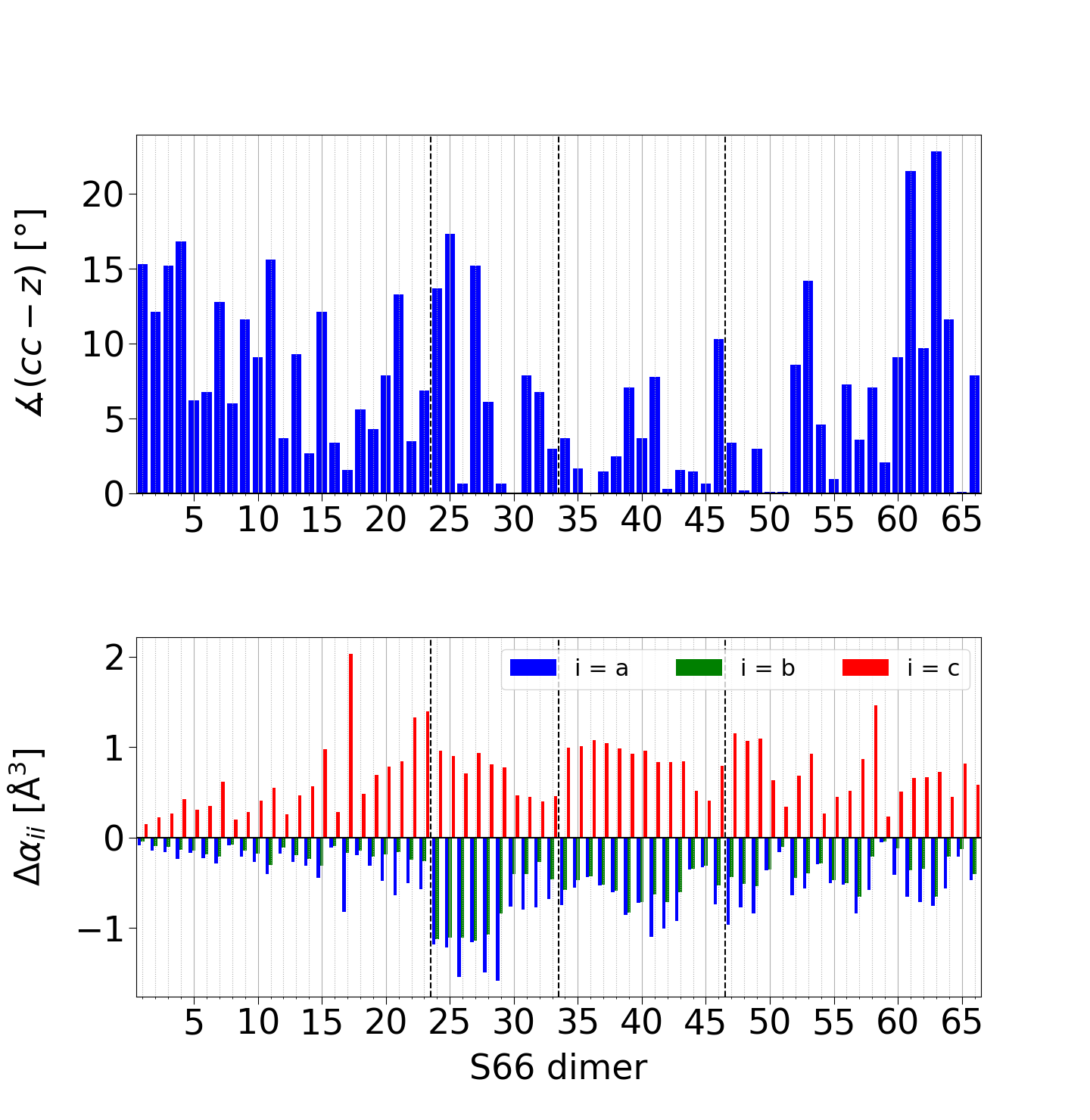}\caption{\label{polar}The angle of the interaction polarizability first principal axis with respect to the \textit{z}-axis as well as the three principal components of the interaction polarizability for the S66 noncovalent dimers calculated with dRPA@PBE0 (``c'' first principal axis, ``b'' second principal axis, ``a'' third principal axis). Vertical black dashed lines separate hydrogen bonding, $\pi$-stacking, London dispersion, and mixed interaction classes.}\end{figure}

The principal axis corresponding to the first principal component includes a small angle (0-22.8\degree{}) with the \textit{z}-axis and is shown in Figure \ref{polar}.
This angle is particularly small for London dispersion bonded dimers (3.3\degree{} on average), while somewhat larger for hydrogen bonded (8.8\degree{} on average) and $\pi$-$\pi$-interaction bonded (7.1\degree{} on average) dimers.
The first principal component ($\boldsymbol{\nu}^c$) of the interaction polarizability has a positive eigenvalue.
The second ($\boldsymbol{\nu}^b$) and third ($\boldsymbol{\nu}^a$) principal components of the interaction polarizability have negative eigenvalues.\\
\indent $\boldsymbol{\nu}^b$ and $\boldsymbol{\nu}^a$ are smaller than $\boldsymbol{\nu}^c$ in the hydrogen bonded cases, indicating the dominance of the directionality imposed by the hydrogen bonds.
In the case of London dispersion and mixed interactions, the relative magnitude of the components depends on the relative orientation of the monomers.
This is in line with our observations on how the monomers' linear and quadratic polarization components strongly determine the interaction polarizability (compare for example dimers 17 and 26, dimers 30 and 40, or dimers 34 and 36 in Figures \ref{mono_pol} and \ref{polar}).

The sign of the interaction polarizability principal components is mostly in line with the intuitive assumption that dimer polarization parallel to the $z$-axis aligns dipolar polarization at opposite poles and increases the intermolecular bonding while dimer polarization in the $xy$-plane aligns dipolar polarization at identical poles and decreases the intermolecular bonding. 
Even though this simple and intuitive design principle for the modulation of the intermolecular interaction via a dipolar field is here confirmed for the S66 set, very strong fields would be required for its successful application.

Figures \ref{selected1} (hydrogen bonded dimers), Figure \ref{selected2} ($\pi$-stacking and London dispersion), and Figure \ref{selected3} (mixed interaction) depict the quadratic dependency of the interaction energy with respect to our employed field strength of up to $\approx$5 GV m$^{-1}$ along the $z$-axis.
It visualizes the range of modulation of the interaction energies up to a moderately strong OEEF.
The range of modulation of the interaction energies by an OEEF in the $xy$-plane can be obtained by inspection of the interaction dipole moment arrows and interaction polarizability ellipsoids in the Figures and also by comparing with Figures \ref{dip} and \ref{polar}.\\
\indent Monomers with more extended $\pi-$conjugations than those present in the S66 set (for example fused rings) generally show a larger charge density polarizability and cooperative polarization coupling via many-body effects\cite{kozlowska,zawada2011} and are therefore capable of larger modulation effects of interaction energies by an OEEF.

\subsection{Intermolecular distance dependence of the interaction induced  properties}
The effect of the OEEF on the noncovalent interaction potential energy curves of the S66 dimers is briefly discussed in this section. 
Table \ref{Distance} depicts values of the averaged interaction energy, dipole moment ($z$-component), and polarizability ($z$-component) discretized for four intermolecular distances $r/r_\mathrm{e}$ ($r_\mathrm{e}$ being the equilibrium distance) along the $z$-axis.
The interaction energy is given for zero field strengths, as well as for our strongest employed field strengths ($\pm \ \approx$ 5.14 10$^{9}$ V m$^{-1}$). 
On average, and in line with the aforementioned concavity with the field, the interaction strengthens upon application of the field in either direction. This observation also holds for the London dispersion dominated dimers (see also Fig.~\ref{selected2}). 
For the hydrogen bonded dimers (shown in Fig.~\ref{selected1}), interaction energies are substantially larger in magnitude, and their  interaction strength can be either decreased or weakened, depending on the field's direction.
As one would expect (due to the cancellation of the nuclear repulsion terms), 
the magnitude of the interaction induced electric properties decreases monotonically with the intermolecular distance.
Not surprisingly, the interaction dipole moment in hydrogen bonded dimers is largest in magnitude and still quite sizeable at twice the equilibrium distance. 
By contrast, interaction dipole moments are negligible for the London dispersion dominated dimers. 
At such a small absolute scale, we think it unlikely that their changing sign for some distances carries 
much meaning due to all numbers in Table~\ref{Distance} corresponding to averages over dozens of chemistries. 
For interaction polarizabilities, an interesting trend is observed: London dispersion dominated dimers 
exhibit significantly larger values than hydrogen bonded dimers for short distances, at twice the
equilibrium distance, however, the interaction polarizability of the hydrogen bonded dimers is larger.
Presumably this is due to London dispersion decaying with a faster power law than electrostatic interactions. 


\begingroup
\begin{table}[htb!]
\caption{\label{Distance} Interaction energies $\Delta E$ [kcal/mol], dipole moments $\Delta \mu$ [D], and polarizabilities $\Delta \alpha$ [\AA$^3$] at CCSD(T)-F12b/aug-cc-pVDZ level for four intermolecular distances, averaged over the entire S66 dimer sets, as well as for dimer sub sets HB and $\pi$ corresponding to being bonded predominantly through hydrogen bonding (shown in Fig.~\ref{selected1}) or London dispersion (shown in Fig.~\ref{selected2}), respectively. 
Averaged interaction energies $\Delta E^n$, $\Delta E^z$, and $\Delta E^p$ correspond to results obtained in electric fields 
with negative, zero, and positive strengths $F_z =$ -0.01, 0.00, and +0.01 a.u., respectively.
 }
\begin{center}
\begin{tabular}{l | c c c c } \hline
$r/r_\mathrm{e}$  & 9/10 & 1  & 5/4 & 2 \\\hline \hline
$\Delta E^n$     & -5.059 & -5.927 & -4.107 & -0.915\\
$\Delta E^z$     & -4.556 & -5.487 & -3.786 & -0.790\\
$\Delta E^p$     & -4.679 & -5.630 & -3.925 & -0.855\\ \hline
$\Delta E^n_{\rm HB}$   & -10.093 & -10.451 & -7.641 & -2.003\\
$\Delta E^z_{\rm HB}$   &  -8.322 &  -8.986 & -6.684 & -1.677\\
$\Delta E^p_{\rm HB}$   &  -7.073 &  -8.024 & -6.158 & -1.555\\ \hline
$\Delta E^n_{\pi}$      & -2.438 & -3.796 & -2.310 & -0.322\\
$\Delta E^z_{\pi}$      & -2.266 & -3.609 & -2.142 & -0.252\\
$\Delta E^p_{\pi}$      & -2.820 & -4.088 & -2.477 & -0.375 \\ \hline \hline
$\Delta \mu$            & 0.356  & 0.292  & 0.189  & 0.067 \\
$\Delta \mu_{\rm HB}$   & 0.823  & 0.671  & 0.427  & 0.147 \\
$\Delta \mu_\pi$        & 0.004  & -0.001 & -0.002 & 0.001 \\ \hline \hline
$\Delta \alpha$         & 0.740  & 0.688  &  0.541 & 0.225 \\
$\Delta \alpha_{\rm HB}$& 0.615  & 0.595  & 0.510  & 0.241 \\ 
$\Delta \alpha_\pi$     & 0.858  & 0.786  & 0.595  & 0.228 \\ \hline \hline
\end{tabular}
\end{center}
\end{table}
\endgroup

\section{Conclusion}

Interaction induced electric properties along the binding axis for the diverse dimers of the S66 set have been presented using high-level coupled cluster computations [CCSD(T)-F12b/aug-cc-pVDZ-F12] as a reference.
Our results suggest that the intermolecular energy's response to variations in the electric field along the intermolecular axis is largely dominated by first (dipole) and second (polarizability) order effects, while higher order contributions (hyperpolarizabilities) play only a negligible role. 
Based on symmetry adapted perturbation theory (SAPT0),
a qualitative assessment has been performed for the most relevant physical interactions that govern the observed interaction induced properties.
Analysis of first order SAPT contributions computed on a monomer-resolved level reveals some essential 
links between composing monomer and dimer properties.
The reference results have served as a benchmark for testing the performance of frequently employed density functional approximations (DFAs) of non-local exchange and correlation in combination with the semi-local PBE exchange-correlation kernel and the aug-cc-pVQZ basis set.
We find that exact exchange contributions play an important role for the performance of the DFAs, raw PBE showing an unsigned relative mean error above 10\%.
Among the computationally more efficient DFAs, the LC-$\omega$PBE0-D3 level of theory stands out providing reliable values for interaction dipole moments and dipole polarizabilities throughout (relative error below 5\%).
A very small relative error of at most a few percent is obtained at the computationally more demanding dRPA@PBE0 level of theory.
The dRPA@PBE0 level of theory has been used to compute the full interaction dipole moment vector, as well as the principal components of the interaction polarizability tensors.
Our tensorial property analysis indicates that the vector of directionality given by hydrogen bonds for manipulating dominates the first order interaction response.
It also reveals that the principal components of the interaction polarizability stay well aligned with the interaction axis in all cases (at most $\approx$20$^\circ$ deviation), and consistently offer one vector for increasing the interaction energy (approximately along the binding axis) and two perpendicular vectors for decreasing the interaction energy.  
Results for intermolecular distance dependent averages of interaction energies, dipole moments, and polarizability also suggest interesting trends for hydrogen bonded vs.~London dispersion bonded systems. In particular, we find that fields can only strengthen interaction energies for London dispersion bonded systems, while they can strengthen {\em and} weaken the hydrogen bonded systems (depending on direction). Interaction dipole moments decay monotonically with distance for hydrogen bonded systems, and are negligible for London dispersion bound systems. Interaction polarizability trends depend on distance: In the equilibrium distance they are larger for London dispersion bound systems, for larger intermolecular distances hydrogen bonded systems exhibit larger interaction polarizabilities. 
We believe that our study of interaction induced electric properties, focusing on accurate reference data, DFA non-local exchange-correlation performance testing, diverse interaction types, and qualitative interpretation of the most relevant physical interactions, bears essential exploratory insights for future endeavours in intermolecular manipulation and design via oriented external electric fields.

\section{Supporting information}

A spreadsheet containing numerical results for all properties discussed in this work is included. 

\section{Acknowledgment}
We acknowledge support from the European Research Council (ERC-CoG grant QML). 
This project has received funding from the European Union's Horizon 2020 research and
innovation programme under Grant Agreement \#772834.
This work was partly supported by the NCCR MARVEL, funded by the Swiss National Science Foundation. 
We also acknowledge support by the Swiss National Science foundation (No.~PP00P2\_138932, 407540\_167186 NFP 75 Big Data, 200021\_175747, NCCR MARVEL).
Some calculations were performed at sciCORE (http://scicore.unibas.ch/) scientific computing core facility at University of Basel.

\bibliographystyle{ieeetr}
\bibliography{references_new}

\end{document}